\documentclass{an}
\usepackage{graphicx}
\usepackage{times}
\Volume{00}              
\Year{0000}              
\Month{00}               
\Pagespan{000}{000}      

\begin{document}
\lhead[\thepage]{K.F. Gunn et al.: Radio observations of the
XMM-Newton/Chandra 13hr deep survey field}
\rhead[Astron. Nachr./AN~{\bf XXX} (200X) X]{\thepage}
\headnote{Astron. Nachr./AN {\bf 32X} (200X) X, XXX--XXX}

\title{Radio observations of the XMM-Newton/Chandra 13hr deep survey
field} 

\author{K.F. Gunn\inst{1},
I.M. M$^{\rm c}$Hardy\inst{1}, N. Seymour\inst{1}, A.M. Newsam\inst{2},
M.J. Page\inst{3}, K.O. Mason\inst{3}, N. Loaring\inst{3},
L.R. Jones\inst{4}, T. Muxlow\inst{5}, T. Takata\inst{6},
K. Sekiguchi\inst{6} \and T. Sasseen\inst{7}} 
\institute{
Department of Physics \& Astronomy, University of
Southampton, Southampton, SO17 1BJ, UK
\and
Astrophysics Research Institute, Liverpool John Moores University,
Birkenhead, CH41 1LD, UK
\and
Mullard Space Science Laboratory, University College London,
Holmbury St Mary, Dorking, RH5 6NT, UK
\and
School of Physics \& Astronomy, University of Birmingham
Edgbaston, Birmingham, B15 2TT, UK
\and
University of Manchester, Nuffield Radio Astronomy Laboratories,
Jodrell Bank, Macclesfield, Cheshire, SK11 9DL, UK
\and
National Astronomical Observatory of Japan, 650 North A`ohoku Place,
Hilo, HI 96720, USA
\and
Department of Physics, University of California, Santa Barbara, 
CA 93106, USA
}
\date{Received {date will be inserted by the editor}; 
accepted {date will be inserted by the editor}} 

\abstract{
Our VLA observations of the XMM-Newton/Chandra 13hr deep survey field
(see Page et al., this proceedings) result in one of the two deepest
1.4GHz radio maps ever made.  Within the $15'$ radius field covered by
the deep X-ray data (0.19 deg$^2$), a total of 556 radio sources are
detected, down to a $4\sigma$ flux limit of $28\mu$Jy.  Of the 214
Chandra sources, 55 have radio counterparts.  The sub-arcsecond
accuracy of the VLA and Chandra positions enable us to determine with
high confidence the sources common to both surveys. Here we present
the relationship between the X-ray and radio source populations at the
faintest radio flux limits yet probed by such a study.  We discuss how
the X-ray/radio relationship differs as a function of optical
morphology, ie between unresolved `stellar' objects and well resolved
galaxies. We then discuss the origin of the X-ray and radio emission,
ie AGN, starburst or a mixture of both, in these two classes of
object.
}
\correspondence{kfg@astro.soton.ac.uk}

\maketitle

\section{Introduction}

Our Chandra/XMM-Newton X-ray survey of the 13hr field ($13^{\rm h}
34^{\rm m} 37^{\rm s} +37^\circ 54' 33''$, J2000) has been undertaken
with the aim of understanding the physics of the X-ray sources around
the knee of the source counts, ie, those sources which produce the
bulk of the X-ray background radiation.  To complement our initial
X-ray survey, we have made a deep VLA 20cm radio observation of the
field, both as a survey in its own right, but also to investigate the
overlap between the faint X-ray and faint radio source populations.
Are they entirely different populations or different manifestations of
the same object?  If the populations overlap, what are the underlying
physics and emission mechanisms in these sources?  How does the
emission mechanism depend upon host galaxy morphology?  (see also, eg,
Bauer et al.~2002).

\subsection{The Data}

The 13hr field was originally a deep ROSAT survey field (100ks with
the PSPC, M$^{\rm c}$Hardy et al.~1998), and was therefore chosen to
be at high galactic latitude in a region of extremely low extinction,
$N_H = 6.5 \times 10^{19} {\rm cm}^{-2}$.  The 13hr field has now been
observed for 200ks with XMM-Newton (Mason et al., in preparation) as
part of the Guaranteed Time programme, and with a mosaic of four 30ks
Chandra ACIS-I pointings (M$^{\rm c}$Hardy et al.~2002) covering 0.19
deg$^2$, reaching $1.3\times 10^{-15}\,{\rm erg\,cm^{-2}\,s}^{-1}$.  A
wealth of supporting data is available for this field, including a
deep Subaru SuprimeCam optical image to $R\sim 27$ and UKIRT/UFTI near
infrared images of the bright X-ray galaxies.  This field will also be
observed at mid to far infrared wavelengths as part of the SIRTF
Guaranteed Time programme, and at 1mm with BOLOCAM.

Since the 13hr field is such a well studied field, it was also chosen
for an extremely deep VLA survey (over 40hrs at 20cm, mostly with the
A-array, plus a further 10hrs at 6cm) reaching $7\mu$Jy rms, making it
one of the three deepest radio surveys ever made (see also Richards et
al.~1998; Ivison et al.~2002).  To study the brightest sources at
higher resolution, 18 days of MERLIN 20cm observations were also
performed.

\subsection{Luminosities}

\begin{figure}
\centerline{
\begin{minipage}{68mm}
\resizebox{\hsize}{!}
{\includegraphics[angle=270]{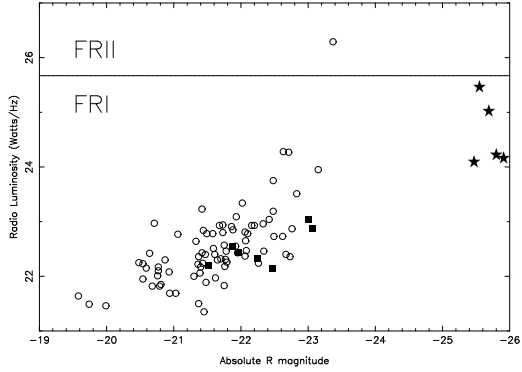}}
\end{minipage}}
\caption{Log radio luminosity {\it vs} $M_R$ for our sub-sample of
radio sources with spectroscopic redshifts ($R<20.5$, 77\% complete),
showing broad line sources, ie, QSOs and BLRGs (filled stars), narrow
line galaxies (open circles), and absorption line galaxies (filled
squares).  The horizontal line at $L_R = 10^{25.6}$ denotes the
conventional distinction between FR II and FR I sources.}
\label{lum}
\end{figure}

Since the VLA survey reaches to such faint fluxes, it is interesting
to look at the typical luminosities for objects in our sample.
Fig.~\ref{lum} shows the radio luminosity {\it vs} optical absolute
magnitude for the sub-sample of sources with $R<20.5$ for which
redshifts are available.  Of these, the majority of radio sources have
$z<1$, and are of relatively low luminosity.  For the fainter sources,
if we assume $z\sim 1.5$, then a $50\mu$Jy source has a luminosity of
$3\times 10^{23}\,{\rm WHz}^{-1}$, which is at the upper limit for
`normal' galaxies; an X-ray source of $10^{-15}\,{\rm
erg\,cm^{-2}\,s}^{-1}$ has $L_X = 6 \times 10^{42}\,{\rm
erg\,s}^{-1}$, at the upper limit for starburst galaxies.

We note also that almost all the radio sources in our sample are
smaller than 50kpc in size, ie, the size of a galaxy or smaller,
consistent with a starburst origin for much of the emission, therefore
we are not detecting the equivalent of classical double-lobed radio
galaxies.

\subsection{Stellarity}

\begin{figure}
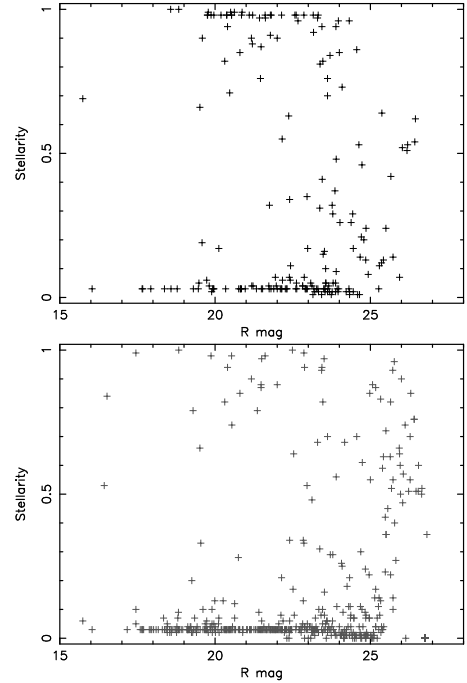

\centerline{
\begin{minipage}{60mm}
\resizebox{\hsize}{!}
{\includegraphics[angle=270]{stellar_chids.ps}}
\resizebox{\hsize}{!}
{\includegraphics[angle=270]{radio_stellarity_rmag.ps}}
\end{minipage}}
\caption{The stellarity of the optical counterparts against their
$R$--band magnitude for both our X-ray (upper) and radio (lower)
samples.  }
\label{stellarity}
\end{figure}

The optical counterparts to the X-ray and radio sources in our samples
are found from the $R\sim27$ Subaru image of the field.  The {\sc
SExtractor} source extraction package (Bertin \& Arnouts 1996) is used
to create the optical catalogue, containing photometric and
morphological information.  The `stellarity' parameter denotes how
resolved an optical source is: broadly speaking, stars have stellarity
$S = 1$ and galaxies $S = 0$, but a crude separation can be made at $S
= 0.5$.  {\sc SExtractor} has difficulty in classifying the morphology
for objects fainter than $R\sim 25$ in our data.  In
Fig.~\ref{stellarity}, we plot the stellarity of the optical
counterparts against $R$--band magnitude, for both the X-ray and radio
samples, which clearly shows that radio counterparts are more
galaxy-like than the X-ray counterparts, as would perhaps be expected
due to orientation effects in the context of the Unified Model of AGN.

\subsection{Optical magnitude distributions}

\begin{figure}
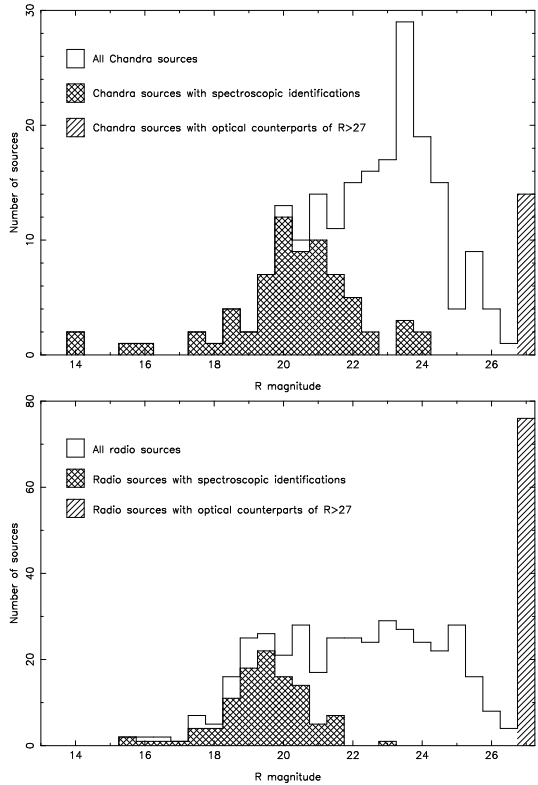

\centerline{
\begin{minipage}{70mm}
\resizebox{\hsize}{!}
{\includegraphics[angle=270]{kfg_rmag.eps}}
\resizebox{\hsize}{!}
{\includegraphics[angle=270]{radio_rmag.eps}}
\end{minipage}}
\caption{The optical magnitude distribution for the X-ray (upper) and
radio (lower) samples.  The shaded areas indicate those sources with
spectroscopic identifications and redshifts.  The drop-off in sources
fainter than $R>25$ is due to catalogue incompleteness and is not a
real effect.  The bin at $R=27$ in each case reflects the sources for
which an optical counterpart is not visible on the Subaru image.}
\label{optmag}
\end{figure}

In Fig.~\ref{optmag}, the optical magnitude distributions of the X-ray
and radio samples are plotted.  The Chandra sample (upper panel) has a
strong peak at $23<R<24$, with very few counterparts with $R>25$.  The
radio sample (lower panel) has a fairly flat magnitude distribution to
$R\sim 25$, and only falls off at $R>25$ due to optical catalogue
incompleteness.  The higher fraction of optically faint sources means
that more remain unidentified at present than in the X-ray sample.
The broad range of optical magnitudes implies no correlation between
radio or X-ray flux and host galaxy brightness.

\section{Radio/X-ray correlation}

\begin{figure}
\centerline{
\begin{minipage}{77mm}
\resizebox{\hsize}{!}
{\includegraphics[]{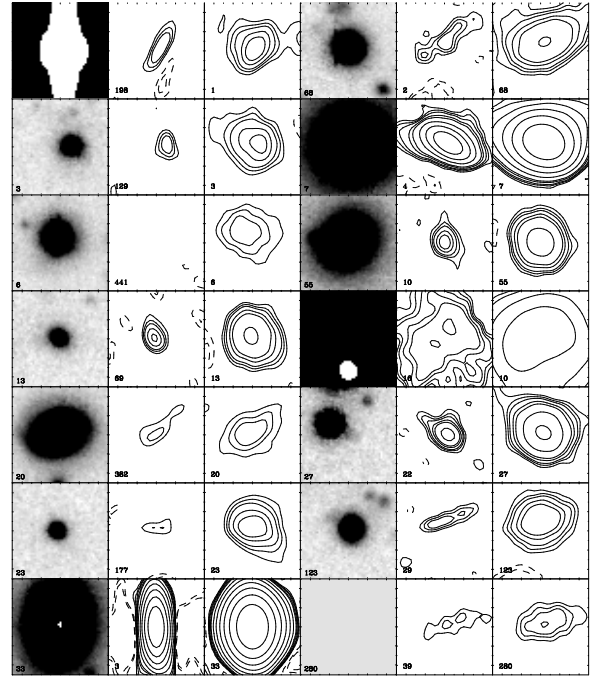}
}
\end{minipage}}
\caption{Optical and radio postage stamp images of the 14 brightest
VLA 20cm sources.  The first panel in each triplet is the optical
image taken from a deep ($R\sim27$) Subaru SuprimeCam exposure.  The
second and third panels are the VLA 20cm data at two different
resolutions, using $1.8''$ and $3.2''$ beams respectively, in order to 
show different aspects of the morphology.}
\label{radx}
\end{figure}

The three coordinate frames, Chandra, Subaru (tied to APM/FK5) and
VLA, are independent and in excellent agreement.  Our source positions
are extremely accurate, Chandra $<0.5''$ and Subaru/VLA $<0.3''$,
enabling confident matching of sources in different wavebands.  Of the
214 Chandra sources and 556 VLA sources in the 13hr field, 55 are
common to both samples.  A wide range of optical magnitudes and
morphologies are found, as shown in Fig.~\ref{radx}, suggesting a
variety of emission mechanisms.

\subsection{Emission mechanisms and diagnostics}

The combination of X-ray, optical and radio data is a powerful tool
for determining the emission mechanism in these faint sources.
Objects expected to be luminous at these wavelengths include normal
broad-line AGN, obscured AGN, BL Lacs, starburst galaxies, and
advective flows, each with distinct properties.  For example,
starburst galaxies are expected to have steep radio spectra and soft,
thermal X-ray spectra, whereas obscured AGN have hard X-ray spectra
and probably flat ($\alpha \sim 0$, $S_\nu \propto \nu^{-\alpha}$)
radio spectra.  Radio galaxies can be either core dominated ($\alpha
\sim 0$) or lobe dominated ($\alpha \sim 0.7$), while BL Lacs have
have a core-halo morphology with flat ($\alpha \sim 0$) radio spectra,
and very steep X-ray spectra.  Advective flows are characterised by a
very flat X-ray spectrum, and probably a higher radio to X-ray flux
ratio than normal AGN. (See eg Condon 1992; di Matteo \& Fabian 1997).

\subsection{Absorption line galaxies}

\begin{figure*}
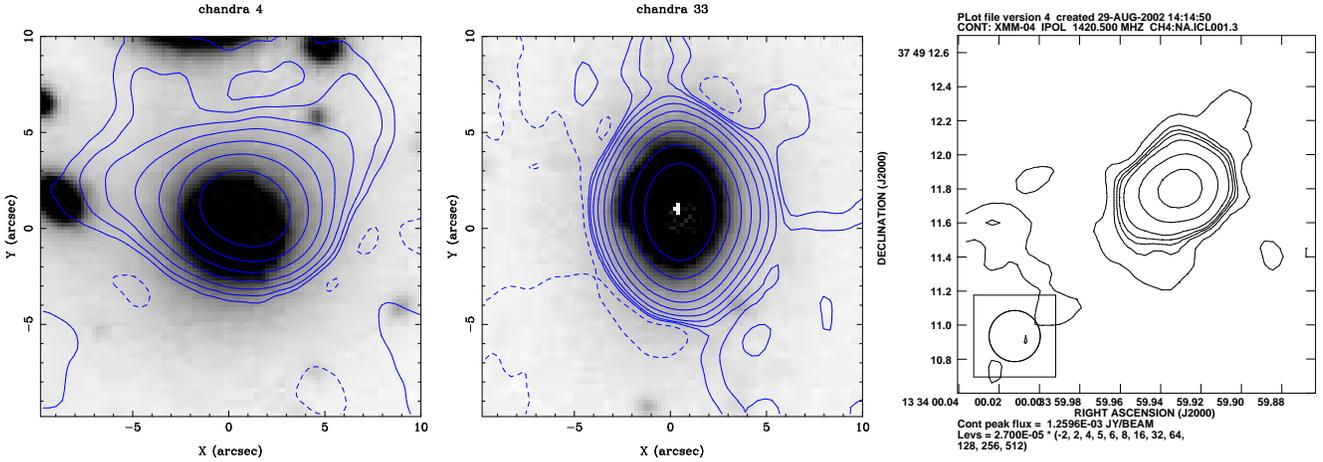

\begin{minipage}{115mm}
\resizebox{\hsize}{!}
{\includegraphics[angle=270]{c4_optrad.ps}
\hspace*{6mm}
\includegraphics[angle=270]{c33_optrad.ps}}
\end{minipage}
\begin{minipage}{60mm}
\resizebox{\hsize}{!}
{\includegraphics[]{c4_merlin.ps}
}
\end{minipage}
\caption{VLA radio contours overlaid on Subaru $R$--band images of
Chandra 4 (left) and Chandra 33 (centre), showing the core-halo
morphology.  Each panel is $20''$ on a side.  A high resolution MERLIN 
map of Chandra 4 is shown in the right hand panel.}
\label{optrad}
\end{figure*}

\begin{figure*}
\begin{minipage}{121mm}
\resizebox{\hsize}{!}
{\includegraphics[angle=270]{c4_xmmspec.ps}
\hspace*{2mm}
\includegraphics[angle=270]{c33_xmmspec.ps}
}
\end{minipage}
\hspace*{10mm}
\begin{minipage}{45mm}
\resizebox{\hsize}{!}
{\includegraphics[]{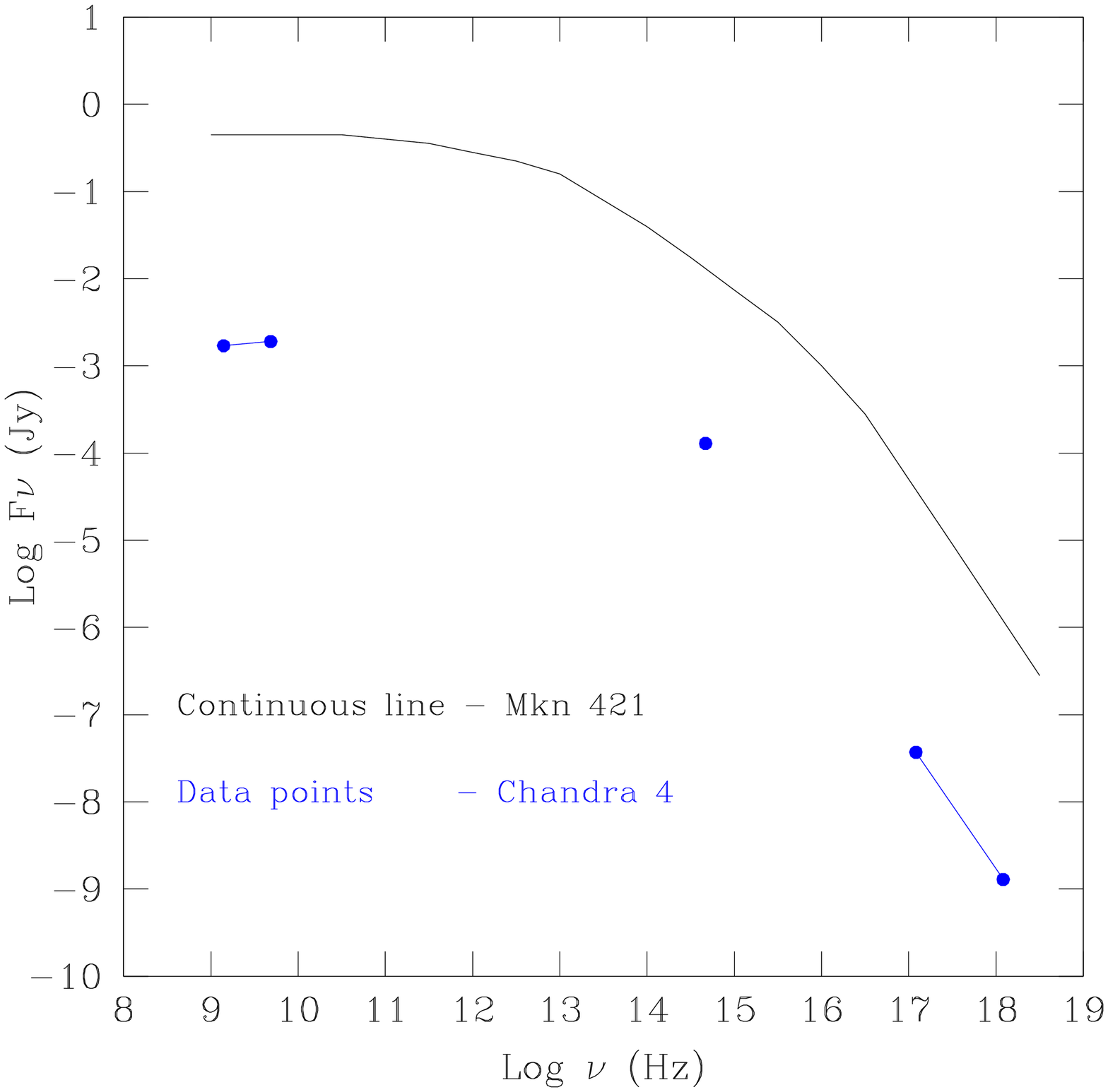}}
\end{minipage}
\caption{The XMM spectra of Chandra 4 (left) and Chandra 33 (centre),
showing steep $\Gamma \sim 2.5$ powerlaw spectra, fixed Galactic $N_H$,
and a small amount of intrinsic absorption at low energies for Chandra
33.  The right hand panel shows the spectral energy distribution of
the well known BL Lac Mkn 421 compared to the available data for
Chandra 4.}
\label{xmmspec}
\end{figure*}

The first deep Chandra survey (Mushotzky et al.~2000) found a
population of optically inactive galaxies with hard spectra, in which
the X-ray emission mechanism is unknown.  Similarly inactive galaxies
were found in our Chandra sample with C4 and C33, the two optically
brightest absorption line galaxies, each having a magnitude of $R\sim
17.5$.  Both are elliptical galaxies with normal optical spectra.
However the radio data showed flat spectra and core-halo morphologies
(Fig.~\ref{optrad}), with very steep X-ray powerlaw spectra
(Fig.~\ref{xmmspec}), indicating that C4 and C33 are almost certainly
BL Lac objects (M$^{\rm c}$Hardy et al., in preparation).
Furthermore, the XMM spectrum of C33 shows a small amount of intrinsic
absorption, $N_H = 1.3 \times 10^{21}{\rm cm}^{-2}$, which could mimic
an intrinsic hard spectrum in a faint source for which only a Chandra
two-band hardness ratio was available.

\subsection{Emission line galaxies}

\begin{figure}
\resizebox{\hsize}{!}
{\includegraphics[angle=270]{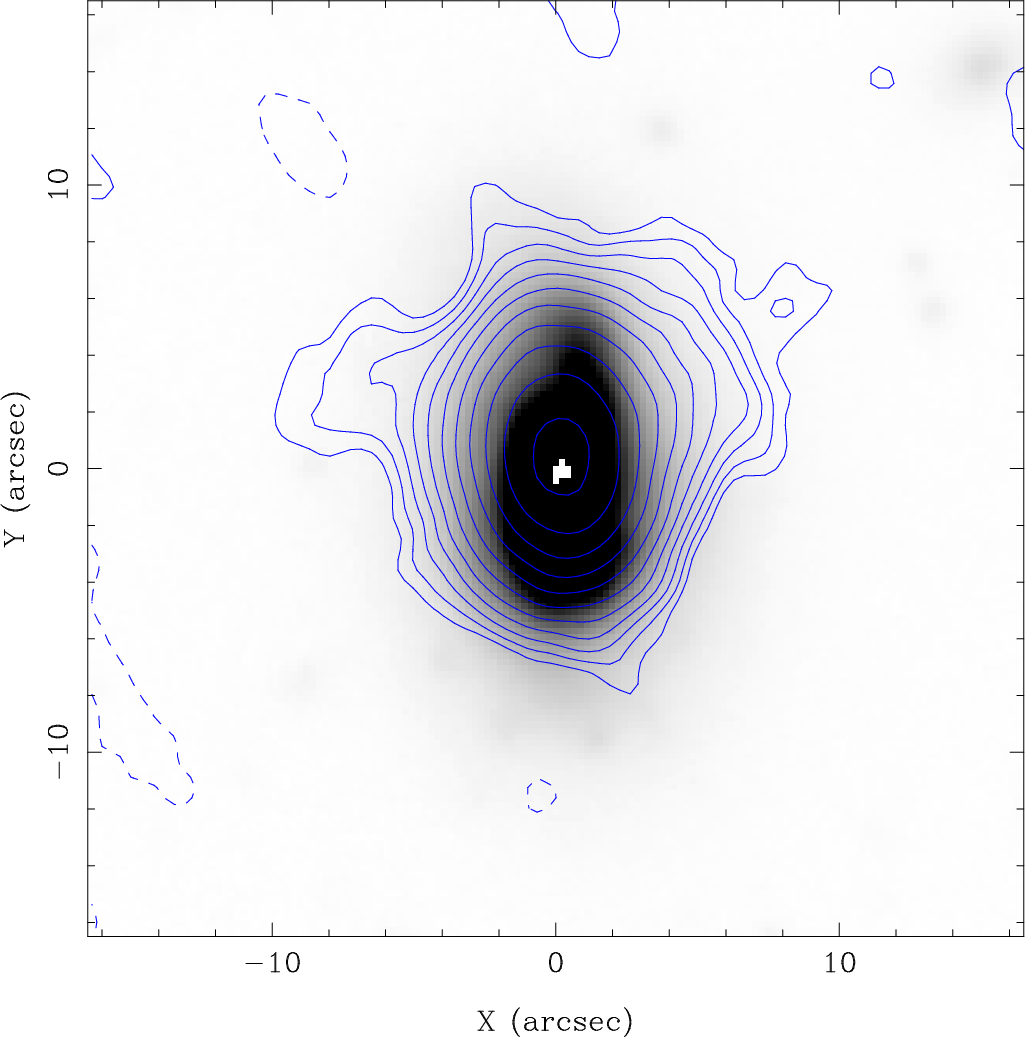}
\includegraphics[angle=270]{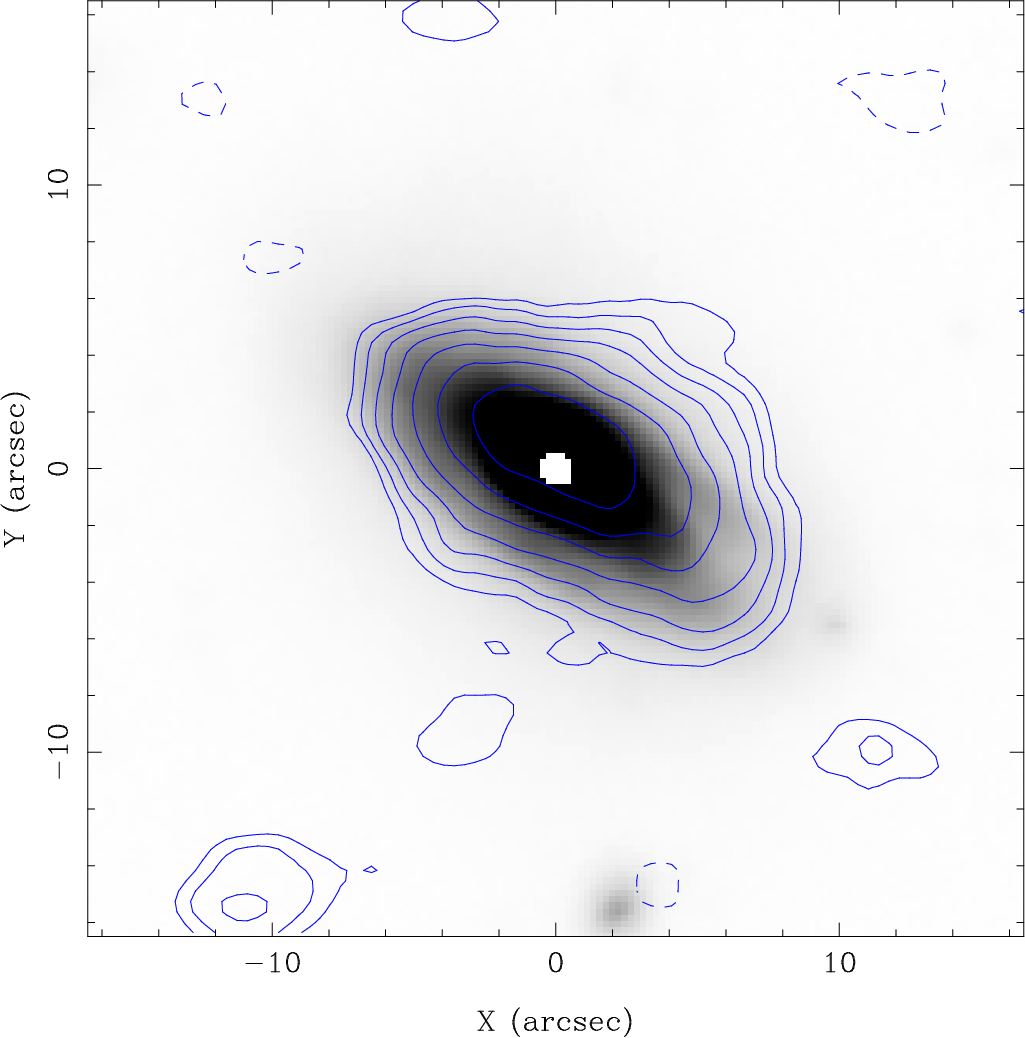}}
\caption{Subaru $R$--band images ($30''$ on a side) of the narrow
emission line galaxies, ROSAT 117 (left) and Chandra 146 (right),
overlaid with VLA A+B array contours ($3.35''$ resolution).  The radio
emission is clearly extended on the scale of the galaxy.}
\label{r32r117}
\end{figure}

Narrow emission line galaxies were found in deep ROSAT X-ray surveys,
with their contribution rising towards faint fluxes (Almaini et
al.~1995; Romero-Colmenero et al.~1996; M$^{\rm c}$Hardy et al.~1998).
The nature of the emission, starburst or obscured AGN activity,
continues to be the subject of debate, but it is becoming apparent
that many sources are a mixture of the two (Page et al., this
proceedings).  It has been suggested that all narrow line galaxies
harbour obscured AGN (Schmidt et al.~1998), however here we describe
two sources for which the X-ray emission is purely due to starburst
activity.  

Fig.~\ref{r32r117} shows $30''$ Subaru $R$--band images of the ROSAT
source R117 (left, Gunn et al.~2001) and the Chandra source C146
(right) with radio contours overlaid.  For both galaxies, which are
optically bright ($R\sim 16$), X-ray faint ($L_X \sim 10^{41}\,{\rm
erg}\,{\rm s}^{-1}$), ie, low $L_X/L_{\rm opt}$, the radio emission
has a steep spectrum ($\alpha \sim 0.6$) tracing the optical extent of
the galaxy, and XMM shows a very soft thermal spectrum, consistent
with a starburst origin.  Furthermore, R117 is not detected by
Chandra, suggesting that the X-ray emission may be extended.  If R117
was a point source, it would have to be a factor of three fainter than
observed with ROSAT to remain undetected by Chandra at this offaxis
angle.  The three optically brightest emission line galaxies in our
sample are starbursts.  Optically fainter emission line galaxies,
however, probably contain an AGN contribution.

\section{Conclusions}

From our VLA data, we obtain radio spectral and morphological
information which, in combination with Chandra and XMM X-ray
data, enable us to show the following results:

\begin{itemize}

\item 55/214 Chandra sources are faint ($>$28$\mu$Jy peak flux) radio 
sources.

\item Faint X-ray and faint radio emission is detected from a variety
of sources, including relatively `normal' galaxies and QSOs.

\item Wide ranges of $L_X/L_{\rm opt}$ and $L_{\rm rad}/L_{\rm opt}$
imply a wide variety of emission mechanisms.

\item At least one third of X-ray/radio matches are normal broad line
AGN. 

\item The two optically brightest absorption line galaxies are almost
certainly BL Lacs.

\item 
The three optically brightest emission line galaxies are
starbursts.

\end{itemize}

\acknowledgements
This work was supported by grants to a number of authors from PPARC
and NASA.  IM$^{\rm c}$H also acknowledges receipt of a PPARC Senior
Research Fellowship.

\end{document}